\documentclass[apjl]{emulateapj} \usepackage{apjfonts}

\shortauthors{Darnley et al.}  \shorttitle{The Angstrom Project Alert
System}

\begin{document}

\title{The Angstrom Project Alert System: real-time detection of
extragalactic microlensing}

\author{M.J.~Darnley\altaffilmark{1}, E.~Kerins\altaffilmark{1,2},
A.~Newsam\altaffilmark{1},
J.P.~Duke\altaffilmark{1}, A.~Gould\altaffilmark{3},
C.~Han\altaffilmark{4}, M.A.~Ibrahimov\altaffilmark{5},
M.~Im\altaffilmark{6}, Y.-B.~Jeon\altaffilmark{7},
R.G.~Karimov\altaffilmark{5}, C.-U.~Lee\altaffilmark{7},
B.-G.~Park\altaffilmark{7} }
\altaffiltext{1}{Astrophysics Research Institute, Liverpool John
Moores University, Twelve Quays House, Birkenhead, Merseyside CH41
1LD, United Kingdom}
\altaffiltext{2}{Jodrell Bank Observatory, University of
Manchester, Macclesfield, Cheshire SK11 9DL, United Kingdom}
\altaffiltext{3}{Department of Astronomy, Ohio
State University, 140 West 18th Avenue, Columbus, OH 43210}
\altaffiltext{4}{Department of Physics, Chungbuk National University,
Chongju 361-763, Korea}
\altaffiltext{5}{Ulugh Beg Astronomical
Institute, Uzbek Academy of Sciences, Tashkent, Uzbekistan}
\altaffiltext{6}{Department of Physics and Astronomy, FPRP, Seoul National
University,
Seoul 151-742, Korea}
\altaffiltext{7}{Korea
Astronomy and Space Science Institute, Hwaam-Dong, Yuseong-Gu, Daejeon
305-348, Korea}


\begin{abstract}
The Angstrom Project is undertaking an optical survey of stellar
microlensing events across the bulge region of the Andromeda Galaxy
(M31) using a distributed network of two-meter class telescopes. The
Angstrom Project Alert System (APAS) has been developed to identify in
real time candidate microlensing and transient events using data from
the Liverpool and Faulkes North robotic telescopes. This is the first
time that real-time microlensing discovery has been attempted outside
of the Milky Way and its satellite galaxies. The APAS is designed to
enable follow-up studies of M31 microlensing systems, including
searches for gas giant planets in M31. Here we describe the APAS and
we present a few example light curves obtained during its commissioning
phase which clearly demonstrate its real-time capability to identify
microlensing candidates as well as other transient sources.
\end{abstract}


\section{Introduction}

Gravitational microlensing has proved itself as a powerful tool for
planetary, stellar and Galactic studies \citep{pac96,beau06,gou06}. It
has also been used to look for dark matter in the Andromeda Galaxy
(M31) where the stellar crowding is much more severe
\citep{cal03,rif03,dej04}. Many of the exciting discoveries made by
microlensing within our own Galaxy are possible only because
microlensing is alerted in real time by survey teams such as OGLE \citep{sumi06}
and MOA \citep{sumi03}, allowing follow-up networks such MicroFUN
\citep{gou06} and PLANET/RoboNet \citep{beau06} to
target candidate events with high time resolution. The requirement of
both rapid and robust data processing has until now prevented the
development of an alert system for M31 microlensing.

The Andromeda Galaxy Stellar Robotic Microlensing (Angstrom) Project
aims to use microlensing to probe the geometry and stellar mass
function of the M31 bulge \citep{ker06}. Since many of these events
are expected to have durations of just a few days, the survey employs a
distributed network of telescopes to enable round-the-clock monitoring
of the M31 bulge. Two of the telescopes within the network are fully
robotic, allowing data to be collected and processed rapidly. The
Angstrom Project is taking advantage of this to develop a real-time
data processing pipeline as part of a microlensing and transients
alert system. The Angstrom Project Alert System (APAS) is designed to
enable follow-up studies of candidate events in M31 in a similar vein
to those conducted within our own Galaxy. \cite{chu06} has highlighted
how such an alert system could enable microlensing discovery of gas
giant planets in M31 itself. Other benefits of such a system would be
a greater ability to identify and characterize binary microlensing
events, which are estimated to comprise around 15\% of all M31 events
\citep{bal01}, and the possibility of measuring finite source effects
which can be used to constrain lens and source parameters.

In this {\em Letter}\/ we describe the design and implementation of
the APAS and we show some early results from its commissioning phase
during the 2006/7 Angstrom observing season. The results illustrate
the potential of the system for real-time microlensing and transient
discovery.

\section{Angstrom Project Observations}

The Angstrom Project\footnote{http://www.astro.livjm.ac.uk/angstrom/}
employs a network of 2m-class telescopes. These include the robotic 2m
Liverpool Telescope (LT) on La~Palma, the robotic 2m Faulkes Telescope
North (FTN) in Hawaii, the 1.8m telescope at Bohyunsan Observatory in
Korea, the 2.4m telescope at MDM in Arizona and, recently, the
1.5m at Maidanak Observatory in Uzbekistan. The telescope network
provides excellent round-the-clock coverage of the M31 bulge during
the observing season, which lasts from August through to
February. Angstrom aims to obtain upwards of three observations per
24-hour period in order to be able to detect and characterize
short-lived microlensing events typically induced by low-mass M dwarf
stars in the M31 bulge \citep{ker06}.

The APAS primarily utilizes incoming data from the two robotic
telescopes, the LT and FTN. These telescopes operate autonomously,
executing programs that are queued by an intelligent scheduler
program \citep{fras06}. Images obtained by these telescopes are automatically
pre-processed and are made available via the web typically within fifteen
minutes of observation. Both telescopes employ identical optical
cameras with a 4.6~arcminute field of view and 2k$\times$2k CCD
arrays. Angstrom observations are conducted primarily in the
Sloan-like $i$ band on both telescopes, typically with a sequence of
exposures totalling around 30~mins. Each exposure within the sequence
is short (no more than 200~secs) in order to avoid saturating the core
of the M31 bulge. The individual exposures are then stacked to form a
single 30~min exposure image for each epoch. More details can be found
in \cite{dar07}, hereafter D07, which describes the main processing
pipeline for the whole telescope network.

In addition to LT and FTN data the APAS uses archive data from the
POINT-AGAPE dark matter microlensing survey of M31
\citep{cal03}. POINT-AGAPE used the wide-field camera of the 2.5m
Isaac Newton Telescope on La~Palma to survey a 0.6~deg$^2$ area of the
M31 disk and bulge. Around 65\% of the LT/FTN reference field overlaps with
the POINT-AGAPE data. For the APAS we have reprocessed all Sloan
$i$-band data from the POINT-AGAPE survey obtained between 1999 and
2001. We use it to provide an extended data baseline for the subset of
light curves lying in both Angstrom and POINT-AGAPE survey regions.

\section{The Angstrom Project Alert System (APAS)}

\begin{figure*}
\epsscale{1.} \plotone{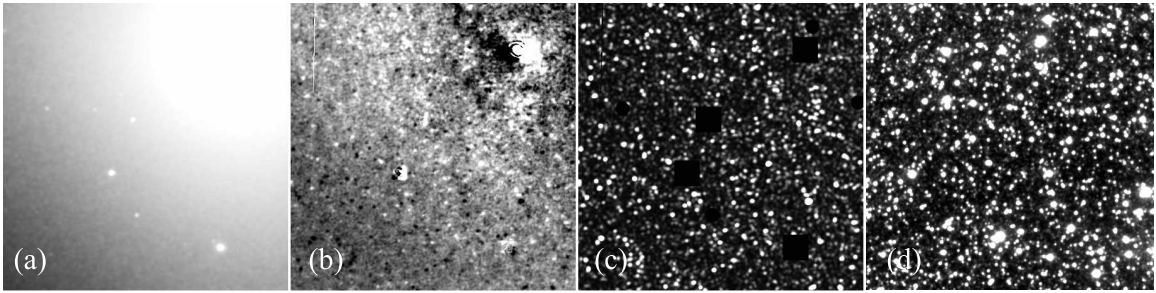}
\caption{(a) LT $i$-band image of a two arcminute region of the Angstrom
M31 bulge field. (b)  Corresponding difference image obtained after
convolving and subtracting a reference image, showing variable objects
as black and white spots. The residual of the imperfectly subtracted
core of the M31 bulge is visible in the top right-hand corner. (c) The
corresponding significance map of variable sources visible above the
background. Variable source blending is evident across the
map. Blackened regions denote masked areas. (d) Stars within a two
arcminute region of an OGLE-III Milky Way bulge field shown for
comparison with the crowding levels seen in the variable source map in
(c). Whilst all objects in (c) have variable flux only a handful of the stars visible in (d) are variables. OGLE image is from the OGLE Early Warning System \citep{uda03}
for alert OGLE-2005-BLG-172. }
\label{images}
\end{figure*}

M31 presents a challenging target for real-time microlensing
discovery. At a distance of 780~kpc it is two orders of magnitude more
distant than typical microlensing sources towards the Galactic bulge
where the majority of Galactic microlensing alerts occur. This means
that larger telescopes and longer exposures are required to obtain
sufficiently sensitive photometry. It also means that we have to deal
with stellar fields which are far more crowded and where the typical
microlensing source is completely unresolved at baseline (see
Figure~\ref{images}). This is the
so-called pixel-lensing regime.

The challenge of developing an alert system for an M31 pixel-lensing
survey differs in some key respects from that of a Galactic
microlensing alert system. One downside is that towards the core of
the M31 bulge we reach the variable star crowding limit. Reliable
photometry for individual variable objects becomes difficult even when
using difference imaging \citep{ala98}. This
means that very often more than one variable source contributes to the
light curve at a given position, as illustrated in the variable source
map shown in panel~(c) of
Figure~\ref{images}. However, variable signals in the M31
bulge must necessarily involve a significant source flux change in
order to be detectable above the very bright M31 bulge surface
brightness. Therefore there are fewer classes of variable stars that
are sufficiently bright and variable to provide potential false
positive signals.

With these potential difficulties in mind, we have adopted a pragmatic
approach in developing the APAS. We do not require it to autonomously
and robustly flag up high quality microlensing alerts. Rather the APAS
presents a modest list of light curves that exhibit transient or
microlensing-like behavior. The list is then reviewed by a human
observer who makes the final decision on whether to issue an alert for
any of the signals. The role of the APAS is to present to the observer
via a web-based interface a compact list (typically no more than a few
dozen light curves) of the currently most interesting transient signals
from a continuously updated database of around 20000 light curves. The
observer is then in a position to quickly review ongoing signals of
potential interest. The APAS also records the positions of all
variable signals in the immediate neighborhood of each light curve in
the list so that the observer can check and quantify any light curve
contamination.

The APAS comprises two stages: a robust real-time data reduction
pipeline and an event classification scheme. We now describe both of these
stages in more detail.

\subsection{Real-time data reduction}
\label {rtdr}

The Angstrom Data Analysis
Pipeline (ADAP) is responsible for reducing all
Angstrom data from both robotic and non-robotic telescopes, and is described in
detail in D07. The nominal mode of operation of the ADAP is to process the
full dataset offline at the end of each observing season. However we
also employ the ADAP to process a 1k$\times$1k re-binned copy of the
2k$\times$2k robotic telescope images that can be processed in real time.
The pixel size of the re-binned data is still only $0.''26$ and so adequately
samples the typical seeing disk ($1''-1.''5$). The ADAP offline processing is
described fully in D07, whereas here we concentrate on the real-time processing
of the re-binned data which forms part of the APAS. Many of these stages are
similar or identical to the ADAP offline processing so we keep our
description brief and refer the reader to D07 for further details.

Firstly, an initial
quality control score is made of each frame based upon the shape and size of
the point-spread function (PSF). Images which fail this check are dropped from
the
subsequent analysis as their inclusion would result in unacceptable
difference image quality. For the surviving images the APAS performs defect
masking, cosmic ray rejection and image alignment. Then the APAS de-fringes each
frame using an iterative method described in D07. Images taken within the same
run (i.e sequentially at one epoch) are then PSF matched and stacked using the
AngstromISIS difference image engine, which is a modified version of a subset of
routines from the ISIS image subtraction package \citep{ala98}. AngstromISIS is
described in detail in D07. The image stack selected to be the reference is then
convolved and subtracted from the other image stacks to create a sequence of
difference images. These are then used to produce one likelihood map of variable
sources for each epoch, using an indicator of statistical significance similar
to that of
\cite{cash79}. Sources from within each likelihood map are then identified and
matched and an updated master list of variable sources is compiled. PSF-fitting
photometry is performed on the difference images and the resulting
photometry is stored within a MySQL database.
The time between observation and adding new
photometric points to the database is about two hours.

This pipeline is
applied both in real time to the LT and FTN data and offline on regions
of legacy POINT-AGAPE data which overlap
with the Angstrom field. The LT and FTN are
essentially identical telescopes and the observations were taken using
matched filter systems, so a simple linear scaling is sufficient to deduce
relative light curve scalings and offsets for these telescopes. The POINT-AGAPE
$i$-band data is matched to LT/FTN $i$-band photometry by fitting a sample of
periodic variables present in both the Angstrom and POINT-AGAPE fields.

\subsection{Candidate classification}

APAS event classification is initiated each time
a new robotic observation is added to the light curve database.  The APAS
employs a
number of selection criteria to classify
each light curve into one of seven different categories.

Each object's light curve is separated by telescope (LT, FTN or POINT-AGAPE),
with each
telescope being initially processed separately. The APAS deems an object to
be valid if it contains at
least seven LT light curve points from the first (2004/5) and second
(2005/6) Angstrom
observing seasons. It must also contain at least seven FTN light curve points
from
the second Angstrom season.  If an object does not meet these criteria
it is classed as {\em Void} and will never be processed again by the
APAS. If a light curve does not contain at least seven
points in the current (2006/7) observing season then the object is marked as
{\em Invalid}. Such objects are reviewed again when new
observations are added to the database.

For remaining objects a baseline is estimated for each telescope by
determining the median light curve flux. The APAS then looks for indications of
transient behavior.  Any point that is at
least $5\sigma$ above the baseline is considered  ``significant''.  For each
object the APAS then investigates any
clusterings of significant points (``peaks'') for each telescope.  A
peak is defined as a clustering of at least five significant points,
all lying within the same observing season.  Within the peak some leeway is
permitted - up to three consecutive non-significant points are admitted
within the peak provided they are followed by a significant point within the
same season and provided their presence coincides with worsening
seeing.

Objects containing no peaks are classed as {\em Flat}.  These
objects will be re-analyzed and possibly re-classified when new observations
are added to the database.
If an object contains no peaks in the current season but contains at
least one peak within a previous
season then it is classed as {\em Variable}, undergoing
re-analysis when additional observations are obtained.
Objects containing peaks in the current
season are checked for the presence of peaks in previous seasons. The
APAS must allow for the fact that the crowding of variables may superpose
variability on the baseline of transient and microlensing events \citep[for example candidate PA-99-N1][]{pau03}. Accordingly,
it does not necessarily classify objects as variable if they show variability
on previous seasons as well as in the current one. Instead, if peaks are seen in
previous seasons then the object is classified as variable unless a peak in the
current
season is at least 150 reference image counts and is at least
50\% greater than the highest amplitude peak in any previous season. Other than
these the only remaining objects are those which show only peaks during the
current season.

The APAS now considers all surviving objects to be transients and
flags them as {\em Followed}.  The FTN and POINT-AGAPE (where available)
light curves for these objects are now scaled to the reference image
of the LT data and a
reduced Paczy\'nski profile is then
fitted with a
flux $F(t)=B+\Delta F/\sqrt{1+12[(t-t_{0})/t_{\rm FWHM}]^2}$,
where $t$ and $t_{0}$ are the epochs of observation and
maximum brightness, $B$ is the baseline flux, $\Delta F = F(t_0)-B$ is the
maximum flux deviation  and $t_{\rm FWHM}$
is the duration of the full-width at half-maximum. The values of
$B$ and $\Delta F$ are telescope and bandpass dependent.
We require that the best fit $t_{0}$
does not occur
before the first observation of the current season and that it does
not occur after the anticipated start of the next season.  The FWHM of
the event must be in the range of $1~\mbox{day} \leq t_{\rm FWHM}\leq365$ days.
Finally we require that the amplitude of the event be
positive.  Objects passing these criteria are
classed as {\em Alerts}.  Objects that fail any of these criteria
remain classed as {\em Followed}, unless they were previously classified as an
alert in which case they are re-classed as {\em Old}.

The APAS utilizes a web-based interface to report results.
Whilst allowing a user to examine the light curves of all alerted
and followed objects it also displays light curves of all
objects within a three arcsecond vicinity of each alert. This is helpful in
deciding whether an object is a true alert but may have a baseline
contaminated by a nearby variable source.

\subsection{Results from the commissioning phase}

\begin{figure*}
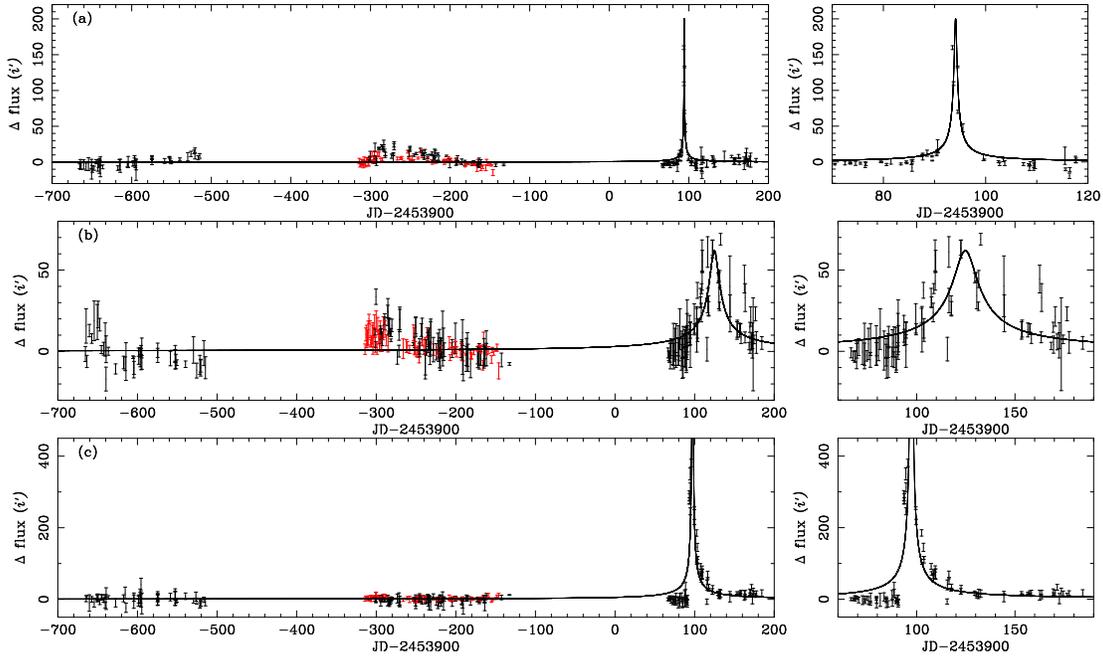

\begin{centering}
\includegraphics[width=2.85cm,angle=270]{f2a}
\includegraphics[width=2.85cm,angle=270]{f2b}\\
\includegraphics[width=2.85cm,angle=270]{f2c}
\includegraphics[width=2.85cm,angle=270]{f2d}\\
\includegraphics[width=2.85cm,angle=270]{f2e}
\includegraphics[width=2.85cm,angle=270]{f2f}\\
\end{centering}
\caption{Three examples of alerts selected by the
 APAS. LT photometry is shown in black and FTN in red. (a) A short
time-scale ($t_{\rm FWHM} \simeq 1$~day) high signal-to-noise ratio microlensing
candidate. (b) A lower
signal-to-noise ratio microlensing candidate at the borderline of identification
by
APAS. Both microlensing candidates exhibit periodic variability in their
baselines due to nearby variable sources. (c) Classical Nova
2006~8\protect\footnote{Rochester naming convention
  http://www.rochesterastronomy.org/snimages/index.html} first
  announced by \cite{atel884}.  This light curve is easily distinguishable
from microlensing by its asymmetric profile. Right-hand panels show zooms of
the events and the solid line is the best-fitting reduced Paczy\'nski curve.}
\label{lc}
\end{figure*}

\begin{table}
\caption{A snapshot statistical summary from the APAS.}
\center{\begin{tabular}{lll}
\hline
\hline
Object & Number of & Percentage\\
status  & objects   & of total \\
\hline
Void     & $6,311$ & $34.0\%$\\
Invalid  & $54$    & $0.3\%$\\
Flat     & $9,931$ & $53.6\%$\\
Variable & $2,197$ & $11.8\%$\\
Follow   & $12$    & $0.1\%$\\
Old      & $0$     & $0.0\%$\\
Alert    & $38$    & $0.2\%$\\
\hline
\end {tabular}}
\label{alert_table}
\end{table}

The Angstrom Project is currently in its third observing
season and this is the first season in which we have implemented the APAS.
Table~\ref{alert_table} contains a snapshot summary of statistics from a
recent run of the APAS. Around $12\%$ of the light curve catalogue
is classified as variable or transient. The bulk of the rest of the
light curves are either ``Flat'' because they do not show significant
light curve variation, or ``Void'' because they do not comprise enough
data points in previous observing seasons. Reassuringly, the number of variables
flagged up by the APAS is in line with the variable number density derived from
an analysis of three years of data from the wide area POINT-AGAPE survey
\citep{an04}. This indicates that, whilst probably not optimal, the APAS
nonetheless competes in terms of sensitivity with previous non real-time
pipelines. We can expect that the offline ADAP analysis of the un-binned data
will be more sensitive still. APAS also succeeds in presenting an easily
manageable number of alerted and followed light curves for a human observer to
review.

Three light curves from the subset of 38 listed alerts are shown in
Figure~\ref{lc}. Clearly, these examples represent credible microlensing and
nova candidates. The APAS therefore demonstrates the capability to provide a
rapid, robust and credible extragalactic alert system. The first candidate in
particular appears to be a very short high signal-to-noise ratio microlensing
event.
Such events tend to be sensitive to finite source effects and this event is
currently the subject of a separate study \citep{duke07}.

\section{Discussion}

We have successfully implemented the Angstrom Project Alert System (APAS), which
performs real-time reduction and identification of extragalactic transient
sources and
microlensing events.  This is the first system of its type to process
extragalactic observations. The APAS has been commissioned during the third
season of Angstrom observations and is already displaying the capability
to identify credible microlensing and nova candidates. During the remainder of
the current 2006/7 observing season and in the following observing seasons we
aim to use the APAS to flag up candidate sources for higher cadence follow-up in
order to obtain detailed light curves of novae outbursts, and to identify
exotic microlensing systems.

\acknowledgments

The LT is operated by Liverpool JMU on behalf of the UK
Particle Physics and Astronomy Research
Council (PPARC).  FTN is operated by the Las
Cumbres Observatory Global Telescope network. Data from FTN was
obtained through the RoboNet-1.0 Consortium Open Time programme. Pre-
processed data from the Isaac Newton Telescope is courtesy of the POINT-AGAPE
team.
EK and JPD are supported, respectively, by an Advanced Fellowship
and a PhD studentship from PPARC.
CH and B-GP are supported by grant C00072 of the Korea Research Foundation.
Y-BJ and C-UL acknowledge support from the Korea Astronomy and Space
Science Institute. MAI and RGK acknowledge the permanent technical and financial support
 of the Maidanak Observatory observations from the Consortium of the
 Korean Universities operating under the MOU between the Consortium and
 UBAI.


\begin{thebibliography}{99}


\bibitem[Alard \& Lupton(1998)]{ala98} Alard, C. \& Lupton, R., 1998, ApJ,
503, 325

\bibitem[An et al.(2004)]{an04}
An, J., et al., 2004, MNRAS, 351, 1071

\bibitem[Baltz \& Gondolo(2001)]{bal01} Baltz, E. \& Gondolo, P., 2001,
ApJ, 559, 41

\bibitem[Beaulieu et al.(2006)]{beau06} Beaulieu, J., et al., 2006,
Nat, 439, 437

\bibitem[Burwitz et al.(2006)]{atel884} Burwitz, V., et al., 2006, The
Astronomer's Telegram, 884, 1

\bibitem[Calchi~Novati et al.(2003)]{cal03} Calchi~Novati, S., et al.,
2003, A\&A, 405, 851

\bibitem[Cash(1979)]{cash79}
Cash, W., 1979, ApJ, 228, 939

\bibitem[Chung et al.(2006)]{chu06} Chung, S.-J., et al., 2006, ApJ,
650, 432

\bibitem[Darnley et al.(2007)]{dar07} Darnley, M.J., et al., 2007, in
preparation (D07)

\bibitem[de~Jong et al.(2004)]{dej04} de~Jong, J.T.A., et al., 2004,
A\&A, 417, 461

\bibitem[Duke et al. (2007)]{duke07} Duke, J.P., et al., 2007, in preparation

\bibitem[Fraser(2006)]{fras06}
Fraser, S., 2006, Astronomische Nachrichten, 327, 779

\bibitem[Gould et al.(2006)]{gou06} Gould, A., et al., 2006, ApJ, 644,
L37

\bibitem[Kerins et al.(2006)]{ker06} Kerins, E., et al., 2006, MNRAS,
365, 1099

\bibitem[Paczy\'nski(1996)]{pac96} Paczy\'nski, B., 1996, ARA\&A, 34,
419

\bibitem[Paulin-Henriksson et al.(2003)]{pau03}
Paulin-Henriksson, S., et al., 2003, A\&A, 405, 15

\bibitem[Riffeser et al.(2003)]{rif03} Riffeser, A., et al., 2003, ApJ,
599, L17

\bibitem[Sumi et al.(2003)]{sumi03}
Sumi, T., et al., 2003, ApJ, 591, 204

\bibitem[Sumi et al.(2006)]{sumi06}
Sumi, T., et al., 2006, ApJ, 636, 240

\bibitem[Udalski(2003)]{uda03} Udalski, A., 2003, AcA, 53, 291
\end{thebibliography}
\end{document}